\documentclass[aps,superscriptaddress,12pt]{revtex4}
\usepackage{amssymb}
\usepackage{amsmath}

\usepackage{graphicx}
\usepackage{dcolumn}

\def\be {\begin{equation}}
\def\ee  {\end{equation}}
\def\bea {\begin{eqnarray}}
\def\eea {\end{eqnarray}}

\begin{document}


\title{Time, vacuum energy,  and the cosmological constant}

\author{Viqar Husain}\email{vhusain@unb.ca}.
\affiliation{Department of Mathematics and Statistics,\\
University of New Brunswick, Fredericton, NB E3B 5A3, Canada}
 

\thispagestyle{empty}

\date{March 30, 2009}

\begin{abstract}
\baselineskip=1.5em

We  describe  a link between
the cosmological constant problem and the problem of time in quantum gravity.    
This arises by examining the relationship  between the cosmological constant
and vacuum energy  in light of non-perturbative formulations of quantum gravity. 
\vfil 

{\small \noindent Written  for the 
Gravity  Research Foundation 2009 Awards for Essays on Gravitation.}
    
\end{abstract}


\maketitle
\eject



\baselineskip=1.5em
Current observations of an accelerating cosmological expansion \cite{expan} have
re-ignited interest in the nature of  dark energy.   One of the putative 
sources of this energy is the cosmological constant  term in Einstein's equation. 
This is needed for inflationary models, and has been hypothesized to arise from
a scalar field in its lowest energy state in the   early universe. 

The cosmological constant is also commonly associated with the vacuum energy 
of quantum fields \cite{revW,revC}.  This  identification is made in the context of the semiclassical 
approximation of quantum fields on a fixed background given by the equation
\be
G_{ab} + \Lambda^0 g_{ab}=  8\pi G\  \langle \Psi|\hat{T}_{ab}|\Psi \rangle,
\label{semiclass}
\ee
where $\Lambda^0$ is a fundamental (or bare) cosmological
constant and $|\Psi\rangle$ is a chosen ``vacuum'' state.  The idea 
that the  expectation value of $\hat{T}_{ab}$ contributes to the observed 
cosmological constant leads to  the identification  of a {\it quantum state dependent effective
cosmological constant}  $\Lambda^{eff}$ given by  
\be 
\Lambda^{eff} := \Lambda^0 - 2\pi G\ \langle \Psi|\hat{T}_{ab}| \Psi \rangle\ g^{ab}.
\label{lamP}
\ee

If the  background is Minkowski  or other highly symmetric spacetime, the matter field vacuum is
more or less unambiguous due to its invariance under the corresponding isometry 
transformations. In such cases it is apparent that 
\be 
  \langle \Psi|\hat{T}_{ab}|\Psi \rangle = \rho g_{ab} 
\ee
holds at least to leading order in $\hbar$, and one can reasonably say that $\rho$ in this equation contributes to the cosmological constant.  

The observed value of the cosmological constant using the cold dark matter model is 
of order $10^{-120}$ in Planck units, whereas the theoretical contribution to it from $\rho$ alone, 
using a suitable cutoff, is in gross disagreement with this number. Indeed, a match between theory and 
experiment requires the finely tuned cancellation of $\Lambda^0$ and $\rho$ to 120 decimal places. 
This is usual statement of the cosmological constant problem. 

Now if the background metric is not highly symmetric and in particular does not have a time like
Killing vector field, the particle concept and vacuum are ambiguous.  This is obviously the case for the large scale FRW universe  we appear to inhabit. So what do we then mean by vacuum energy, and should the assumptions  underlying the statement of cosmological constant problem be revisited?

We argue that the  mantra-like association of the oscillator vacuum energy sum with the cosmological constant must be reexamined in a non-perturbative context, and that doing so provides some insight into whether the problem really exists at a more fundamental level, outside  the semiclassical approximation.
The problem is often viewed in more intuitive terms as the question of whether the matter vacuum gravitates: matter virtual pair production gives an energy density  
\be 
  \rho_{\rm pair} = \frac{2m}{(\hbar/mc)^3} \sim m^4,
\ee 
which is enhanced  to 
\be 
\rho_{\rm pair-graviton\ exchange} = \frac{Gm^2}{\hbar/mc} \  \frac{1}{(\hbar/mc)^3} \sim m^6. 
\ee
if there is a gravitational interaction between pairs: This suggests that the matter vacuum is unstable in the semiclassical approximation, which is another aspect of the cosmological constant problem.

Our first observation is that we ought really to be seeking a vacuum (or ground state) of a full non-perturbativematter-geometry Hamiltonian derived from general relativity coupled to matter fields. In the canonical approach to quantum gravity, the  central problem is to properly formulate and solve the Dirac quantization condition,  which  is the  Wheeler-deWitt equation. Associated with this approach is the problem of time \cite{kuchar-time}: this equation does not look like a time dependent Schrodinger equation because the Hamiltonian constraint of general relativity is not linear in any momentum variable,  and cannot be made so \cite{torre} without adding special matter reference frames \cite{bk}.

Finding a non-vanishing Hamiltonian in the context of any  generally covariant theory therefore requires 
a solution to the problem of time in quantum gravity.  No general solution to this problem is known,  but it is often bypassed by simply making a choice of time gauge by using a phase space variable as a clock.  This process leads to a nonvanishing matter-geometry Hamiltonian density that depends  explicitly on the clock variable for all but the simplest models:
\be 
 \hat{H}= \hat{H}(\hat{q},\hat{\pi};
\hat{\phi},\hat{P}_\phi; \Lambda^{0},g_i;t).
\ee
It also  depends on  the remaining (non-clock) canonical gravity $(\hat{q}_{ab},\hat{\pi}^{ab})$ and matter $(\hat{\phi},\hat{P}_\phi)$
operators, the cosmological constant, and other coupling constants
$g_i$ that may appear in the matter terms. At this stage the $\Lambda^0$  should be viewed as just another parameter in the  Hamiltonian, with no direct connection with vacuum energy. 

Having determined a non-vanishing Hamiltonian, the task is to find its  ground 
state(s) $|q,\phi\rangle_0$  and compute the ground state (or vacuum) energy.  It is at this stage
that there may be an emergent "vacuum energy problem" if
the energy of the relevant state of $\hat{H}$ does not match the
observed one, ie.   if it turns out that  satisfying the equation
\be
\ _0\langle q,\phi|\ \hat{H}  |q,\phi\rangle_0 \equiv
\rho_0(\Lambda^{0},g_i;t) \sim \rho^{(obs)}
\label{qgcosm}
\ee
requires fine tuning of $\Lambda^{0}$ when the present value 
of time $t$ and observed values of the coupling constants $g_i$ 
are inserted  in $\rho_0$.
Furthermore, since the expectation value has explicit time
dependence, it is evident that to agree with observations, the
observed value of vacuum energy density must be time varying.

What is apparent from these observations is that if one starts
from a background independent gravity-matter theory,  either the problem
of time must be solved, or a suitable clock must be chosen,  before one
can even ask if there is a vacuum energy  problem. Furthermore the 
connection of this energy with cosmological constant is  not what is 
suggested by the semiclassical equation (\ref{semiclass}), but is more involved 
with an element of explicit time dependence as suggested by eqn. (\ref{qgcosm}).

For a concrete illustration of this line of argument, let us consider the Hamiltonian of FRW 
cosmology coupled to a scalar field. The phase space variables are the canonical
pairs $(a,p_a)$ and $(\phi, p_\phi)$. The Hubble time gauge in the canonical theory 
is obtained by setting $t=a^2/p_a$ \cite{hw-sc}. The corresponding time  dependent 
Hamiltonian is  given by 
\be 
H^2  = \frac{8\pi G p_\phi^2}{6t^2\left[ 3/8 - 3t^2(\Lambda^0 + 8\pi G V(\phi)\right]},
\ee
where $V(\phi)$ is the scalar field potential.  If $V$ does not vanish, quantization of the 
Hamiltonian is not straightforward; if it does, the eigenstates are those of a free particle with a time 
and cosmological constant dependent mass.  Thus we see that even in this simplest of systems, the connection between the ground state of the Hamiltonian  and the cosmological constant will not be simple. An explicit example of such a relation arises for the case $V(\phi)=0$ \cite{andreas}. 
Furthermore a different time choice will lead to a different Hamiltonian, and a different 
realization of eqn. (\ref{qgcosm}).  
 
In summary, we have demonstrated that at the non-perturbative level there is a  relationship
between the cosmological constant, time  and vacuum energy which is rather complex, and 
fundamentally  different from  what one would conclude from  a naive use of the semiclassical 
equation.  The link between these quantities suggests that at the very least the statement of the cosmological constant problem should be reformulated, and that it may turn out not to be
a problem at all from a non-perturbative perspective.

\medskip

{\it Acknowldgements} This work was supported in part by NSERC of Canada, and the Atlantic Association for Research in the Mathematical Sciences (AARMS).

  \end{document}